\documentclass[epj]{svjour}          

\RequirePackage[T1]{fontenc}

\smartqed  

\usepackage{amsmath, amssymb, bm, graphicx, graphics, color, mathrsfs}
\usepackage{multirow}
\usepackage{amsmath}
\RequirePackage{graphicx}
\RequirePackage{mathptmx}      
\RequirePackage{flushend}
\RequirePackage[numbers,sort&compress]{natbib}
\RequirePackage[colorlinks,citecolor=blue,urlcolor=blue,linkcolor=blue]{hyperref}

\journalname{Eur. Phys. J. C}

\begin{document}

\title{Polytropic stars in Palatini gravity}
\author{Aneta Wojnar
\thanks{\emph{e-mail:} aneta.wojnar@poczta.umcs.lublin.pl}%
}                     
\offprints{}          
\institute{Institute of Physics,
Maria Curie-Sk{\l}odowska University \\
20-031 Lublin, pl. Marii Curie-Sk{\l}odowskiej 1, Poland\label{addr1}\\
and\\
N{\'u}cleo Cosmo-ufes \& PPGCosmo, Universidade Federal do Esp{\'i}rito Santo,\\
29075-910, Vit{\'o}ria, ES, Brasil.
}
\date{Received: date / Revised version: date}
%
\abstract{
We have derived a modified Lane-Emden equation for the Starobinsky model in Palatini gravity which is numerically solvable. Comparing the results to the ones provided by
General Relativity we observe a significant difference depending on the theory parameter for the $M-R$ relations. } 

\maketitle
\date{Received: date / Accepted: date}
\newcommand  {\Rbar} {{\mbox{\rm$\mbox{I}\+!\mbox{R}$}}}
\newcommand  {\Hbar} {{\mbox{\rm$\mbox{I}\!\mbox{H}$}}}
\newcommand {\Cbar}{\mathord{\setlength{\unitlength}{1em}
     \begin{picture}(0.6,0.7)(-0.1,0) \put(-0.1,0){\rm C}
        \thicklines \put(0.2,0.05){\line(0,1){0.55}}\end {picture}}}
\newcommand{\be}{\begin{equation}}
\newcommand{\ee}{\end{equation}}
\newcommand{\ben}{\begin{eqnarray}}
\newcommand{\een}{\end{eqnarray}}
\section{Introduction}
\label{sec:introduction}
The shortcomings \cite{Copeland:2006wr,Nojiri:2006ri,Capozziello:2007ec,Carroll:2004de,Sotiriou:2008ve} of General
Relativity (GR) \cite{Einstein:1915ca,Einstein:1916vd} make the search for some other proposals describing the gravitational phenomena 
necessary and appealing. The dark matter idea \cite{Cap_Laur, Cap_Far}, inflation \cite{Starobinsky:1980te,Guth:1980zm}, 
the fact of the late-time cosmic acceleration \cite{Huterer:1998qv, Sami} with an explanation in the form of the exotic fluid called dark energy 
\cite{Copeland:2006wr,Nojiri:2006ri,Capozziello:2007ec} are just the most widespread problems which we face. Looking for a generalization of Einstein's theory 
is additionally supported by the fact that GR is non-renormalized while adding extra high curvature 
terms seems to improve the situation \cite{stelle}. This is why Extended Theories of Gravity (ETG) \cite{Cap_beyond, cap_invar}
have gained a lot of attention. However, many extensions introduce ghost-like instability. Nonetheless, attacking the Hilbert-Einstein action 
appears in many different ways: assumption on the ``non-constancy'' of the Nature constants
 \cite{Dabrowski:2012eb, Leszczynska:2014xba, Salzano:2016pny}, minimally or non-minimally coupled scalar fields added to the Lagrangian
\cite{brans, Bergmann}, or more complicated functionals than the simple linear one used in GR, for example $f(R)$
gravity \cite{buchdahl, Starobinsky:1980te}. The extra geometric terms coming from the latter approach could explain not only dark matter
issue \cite{cap_jcap, cap_not} but also the dark energy problem. Since the field equations also differ from the Einstein's ones, they usually provide different
behavior of the early Universe. 
One can formulate $f(R)$ gravity in different ways: in the metric approach,
\cite{Sotiriou:2008rp,Carroll:2004de,Sotiriou:2008ve,DeFelice:2010aj,Will:1993te}
Palatini one \cite{Palatini:1919di,Sotiriou:2008rp,Capozziello:2011et,Ferraris:1992dx} as well as hybrid \cite{hybrid}.
We will focus on the Palatini approach in this work.

 The Palatini approach provides modified Friedmann equations \cite{alle_bor1, alle_bor2, alle_bor3} that can be compared with the observational
data \cite{bor_kam, BSSW1, BSSW2, BSS, SSB, wojn_galax}. It shows the potential of the Palatini formulation and it is still applied to gravitational 
problems \cite{lavinia, roshan}. Moreover, there have appeared possibilities of observable effects in microscopic systems providing constraints on models parameters 
\cite{ronco,Lobo:2014nwa}.
There are also disadvantages reported: lack of perturbative approach \cite{Flanagan:2003rb}, conflict with the Standard Model of particle physics 
\cite{Iglesias:2007nv, Olmo:2008ye}, 
the algebraic dependence of the post-Newtonian
metric on the density \cite{Olmo:2005zr, Sotiriou:2005xe}, and the complications with the initial values problem in the 
presence of matter \cite{Ferraris:1992dx, Sotiriou:2006hs}, although 
that issue was already solved in \cite{olmo_sanch}. A similar discussion was performed in 
\cite{vignolo} where it was shown that the initial value problem is well-formulated in presence of the standard matter sources while
the well-posedness of the Cauchy problem should be considered case by case: the Starobinsky one, which we are interested in, belongs to the well-posed class of models.
There are also additional arguments showing that
treating the extra terms in the fluid-like manner provides limitations \cite{zaeem, zaeem2, zaeem3}. However,
as it was shown in \cite{olmo_tri}, higher curvature corrections do not cause the above mentioned problems. Interestingly, the effective dynamics of Loop Quantum
Gravity as well as brane-world cosmological background histories may be reproduced by the Palatini gravity giving the link to one of the approaches to
Quantum Gravity \cite{olmo_singh, olmo_brane}.

There are also astrophysical aspects of Palatini gravity, for instance black 
holes were considered in \cite{diego1,diego2,diego3,diego4,diego5,diego6}, and also wormholes \cite{worm,diego7,diego8} and neutron stars \cite{kain,reij,pano,anab, anet}.
Our concern is related to the last objects in Palatini gravity, especially that the recent 
neutron stars' merger observation \cite{gw} will provide the possible confrontation of gravitational theories. Neutron stars seem to be perfect objects for 
testing theories at high density regimes: there are claims that
using General Relativity in the case of strong gravitational fields and in the case of large spacetime curvature \cite{controversialmag, eksi, berti} is an 
extrapolation.

Stars in Palatini gravity were considered in \cite{Barausse:2007ys, barau, barau2, pani, sham} where it was claimed that there exist surface singularities 
of static spherically symmetric objects in the case of polytropic equation of state which can lead to infinite tidal forces on the star's surface. An argument against that 
claim was introduced in \cite{olmo_ns}: the problem is caused by the particular equation of state which should not be used at the surface. Moreover,
in \cite{fatibene} it was indicated that the polytropic equation of state is nothing fundamental but rather an approximation of the matter forming a star. The another 
important point was mentioned that Palatini gravity should be interpreted according to the Ehlers-Pirani-Schild (EPS) approach
\cite{eps, mauro, fatibene1}, which we are going to follow in this work. That means that a conformal metric is the one responsible for the free fall 
in comparison to metric which was used in \cite{Barausse:2007ys}. As shown in \cite{fatibene}, in this case the singularities are not generated and polytropic stars can be 
obtained in the Palatini framework.

Using this result, we are going to study non-relativistic stars with the polytropic equation of state in $f(\hat{R})$ Palatini gravity. As a working example we will
 use the Starobinsky model, that is, $f(\hat{R})=\hat{R}+\beta\hat{R}^2$ \cite{Starobinsky:1980te}. In order to do it for the
wide class of the stellar objects, we will write down the modified 
Lane-Emden equation obtained from the generalized Tolman-Oppenheimer-Volkoff (TOV) equation which we studied in the context of star's stability in \cite{anet}. That will allow
to examine further different types of stars since the equations describing them can be solved numerically. There are already works considering modified TOV equations
\cite{AltNS1, AltNS2, AltNS3, AltNS4, AltNS5, AltNS6, AltNS7, aw1, aw2} as well as ones providing generalization of the Lane-Emden
\cite{riazi, capH, saito, andre, sak, koy}.
 
We are using the Weinberg's \cite{weinberg} signature convention, that is, $(-,+,+,+)$, with $\kappa=-8\pi G/c^4$.


\section{Stellar objects in Palatini gravity}

\subsection{Palatini formalism}
Before we focus on the Palatini gravity itself, let us briefly present the idea of EPS interpretation \cite{EPS} already mentioned in the introduction. We are interested in 
the formalism in the way as it was presented in \cite{Fatibene, felicia}.
Thus, the formalism assumes that spacetime geometry might be described by two structures, that is, conformal and projective ones. The conformal one concerns a class of
Lorentzian metrics related to each other by the conformal transformation of the form
\begin{equation}
 h=\Omega g,
\end{equation}
where $\Omega$ is a positive defined function. One often interprets such a transformation as a change of frame, e.g. as it is performed in scalar-tensor theories.
On the other hand, the projective structure is a class of connections satisfying
\begin{equation}\label{connec}
 \tilde{\Gamma}^\alpha_{\beta\mu}=\Gamma^\alpha_{\beta\mu} +A_{(\mu}\delta^\alpha_{\beta)},
\end{equation}
with $A_\mu$ is a $1$-form. Thanks to the assumption on the positivity of the conformal function $\Omega$, the conformal structure defines light cones as well as 
provides timelike, lightlike and spacelike direction in a given spacetime. It will determine lengths of timelike and spacelike curves
if one chooses a representative of the conformal class. Moreover, geodesics in a spacetime are defined by a connection. Different connections which belongs to some
projective structure define the same geodesics but parametrized in two different ways \cite{Fatibene}.
A parametrization is a subject of clocks choice, that is, metrics.

The two structures are EPS-compatible if
\begin{equation}\label{condition}
 \tilde{\nabla}_\mu g_{\alpha\beta}=2A_\mu g_{\alpha\beta}.
\end{equation}
It means that if there exists $A$ such that the above equation (\ref{condition}) is true for a metric $g$, then there exists a $1$-form $\tilde{A}$ for any other metric
$\tilde{g}$ such that (\ref{condition}) selects the same connection $\tilde{\Gamma}$. When we consider a triple which consists of the spacetime manifold $M$ and
EPS-compatible structures on $M$, then we deal with EPS geometry. We may put some extra conditions on it: when $A$ is fixed in the way that it depends
on $g$ which had been chosen in the conformal structure than the geometry is called a Weyl geometry. It is integrable when there exists a
connection $\tilde{\Gamma}$ which is a Levi-Civita connection of $\tilde{g}$. In that case there exists a relation between $A$ and the conformal factor $\Omega$:
$A_\mu=\partial_\mu\Omega$ \cite{Fatibene}.

It should be noticed that GR is a special case of the EPS formalism where it is assumed that the connection
$\tilde{\Gamma}$ is a Levi-Civita connection of the metric $g$ (the $1$-form $A$ is zero). Thus one treats the action of the theory as just metric-dependent. But
we may consider the 
Einstein-Hilbert action which depends on two independent objects: the metric $g$ and the connection $\tilde{\Gamma}$. This approach
is called Palatini formalism. Since one uses the simplest gravitational Lagrangian which is linear in scalar curvature $R$, Palatini approach 
turns out to provide that $\tilde{\Gamma}$ is
a Levi-Civita connection of the metric $g$. The difference is that this is the dynamical result, not a assumption as it happens in the previous case.
However, the situation is different when we are interested in more complicated Lagrangians like the ones appearing in ETGs.

Therefore, coming back to the $f(\hat{R})$ Palatini gravity, we see that its geometry is characterized by two independent structures: the metric $g$ and the 
connection $\hat{\Gamma}$. From the field equations it turns out that the connection is a Levi-Civita connection of 
a metric conformally related to $g$. Therefore, one should consider motion of a mass particle provided by the geodesic equation with the connection $\hat{\Gamma}$. 
Clocks and distances in contrast are measured by the metric $g$.
Thus, we are supplied with the action
\begin{equation}
S=S_{\text{g}}+S_{\text{m}}=\frac{1}{2\kappa}\int \sqrt{-g}f(\hat{R}) d^4 x+S_{\text{m}}(g_{\mu\nu},\psi_m),\label{action}
\end{equation}
where $\hat{R}=\hat{R}^{\mu\nu}g_{\mu\nu}$ is the Palatini-Ricci scalar. The variation of (\ref{action}) with respect to the metric $g_{\mu\nu}$ gives
\begin{equation}
f'(\hat{R})\hat{R}_{\mu\nu}-\frac{1}{2}f(\hat{R})g_{\mu\nu}=\kappa T_{\mu\nu},\label{structural}
\end{equation}
where $T_{\mu\nu}$ is energy 
momentum tensor which later on we will assume to be a perfect fluid one while primes denote derivatives with respect to the function's argument: $f'(\hat{R})=\frac{df}{d\hat{R}}$.
On the other hand, the variation with respect to the independent connection provides
\begin{equation}
\hat{\nabla}_\beta(\sqrt{-g}f'(\hat{R})g^{\mu\nu})=0,\label{con}
\end{equation}
from which we immediately notice that $\hat{\nabla}_\beta$ is the covariant derivative calculated with respect to $\Gamma$, that is, it is the Levi-Civita connection 
of the conformal metric
\begin{equation}\label{met}
h_{\mu\nu}=f'(\hat{R})g_{\mu\nu}.
\end{equation}
Moreover, the trace of the equation (\ref{structural}) with respect to $g_{\mu\nu}$ gives rise to
\begin{equation}
f'(\hat{R})\hat{R}-2 f(\hat{R})=\kappa T\label{struc}
\end{equation}
which is called a structural equation. Here, $T$ is the trace of the energy-momentum tensor.
If it is possible to solve (\ref{struc}) as $\hat{R}=\hat{R}(T)$ we observe that $f(\hat{R})$ is also a function
of the trace of the energy momentum tensor, where $T=g^{\mu\nu}T_{\mu\nu}\equiv 3p-c^2\rho$.

 It can be shown \cite{DeFelice:2010aj} that one may rewrite the field equations as a dynamical equation for the conformal metric $ h_{\mu\nu}$ \cite{BSS,SSB} and the scalar 
 field defined as $\Phi=f'(\hat{R})$:
 \begin{subequations}
	\begin{align}
	\label{EOM_P1}
	 \bar R_{\mu\nu} - \frac{1}{2} h_{\mu\nu} \bar R  &  =\kappa \bar T_{\mu\nu}-{1\over 2} h_{\mu\nu} \bar U(\Phi)
	\end{align}
	\begin{align}
	\label{EOM_scalar_field_P1}
	  \Phi\bar R &  -  (\Phi^2\,\bar U^(\Phi))^\prime =0
	\end{align}
\end{subequations}
where we have introduced $\bar U(\Phi)=\frac{\hat{R}\Phi-f(\hat{R})}{\Phi^2}$ and appropriate energy momentum tensor $\bar T_{\mu\nu}=\Phi^{-1}T_{\mu\nu}$.
One also bears in mind that $\hat R_{\mu\nu}=\bar R_{\mu\nu}, \bar R= h^{\mu\nu}\bar R_{\mu\nu}=\Phi^{-1} \hat R$ and $h_{\mu\nu}\bar R=\ g_{\mu\nu}\hat R$.
The last equation, together with the trace of (\ref{EOM_P1}), can be replaced by
\begin{equation}\label{EOM_P1c}
 \Phi\,\bar U^\prime(\Phi)  + \kappa \bar T = 0\,.
\end{equation}
Thus the system (\ref{EOM_P1}) - (\ref{EOM_P1c}) corresponds to a scalar-tensor action for the metric $h_{\mu\nu }$ and (non-dynamical) scalar field $\Phi$
\begin{equation}\label{action2}
 S(h_{\mu\nu},\Phi)=\frac{1}{2\kappa}\int\mathrm{d}^4x\sqrt{-h}\bigg(\bar R- \bar U(\Phi) \bigg) + S_m(\Phi^{-1}h_{\mu\nu},\psi),
\end{equation}
where
\begin{equation}\label{em_2}
    \bar T^{\mu\nu} =
-\frac{2}{\sqrt{-h}} \frac{\delta}{\delta h_{\mu\nu}}S_m  = (\bar\rho+\bar p)\bar u^{\mu}\bar u^{\nu}+ \bar ph^{\mu\nu}=\Phi^{-3}T^{\mu\nu}~,
\end{equation}
and $\bar u^\mu=\Phi^{-{1\over 2}}u^\mu$, $\bar\rho=\Phi^{-2}\rho,\ \bar p=\Phi^{-2}p$, $\bar T_{\mu\nu}= \Phi^{-1}T_{\mu\nu}, \ \bar T= \Phi^{-2} T$ (see e.g. \cite{DGB}).
Further, the trace of (\ref{EOM_P1}), provides
\begin{equation}\label{EOM_metric_2}
    \bar R= 2\bar U(\Phi)-\kappa \bar T.
\end{equation}

\subsection{Generalized Tolman-Oppenheimer-Volkoff equation}
Many models of modified gravity provides changes to TOV equations which influence its macroscopic characteristics, that is, mass and radius of the star when we are 
supplied with an equation of state.
Due to that fact, 
it is possible to introduce some parameters to the equation \cite{aw1} which allow to observe the modifications' influence and classify theories \cite{aw1, szwab}.
An interesting class of modified field equations are the one for which we may write \cite{mim, mim2, mim3}
\begin{equation}\label{mod1}
 \sigma(\Psi^i)(G_{\mu\nu}-W_{\mu\nu})=\kappa T_{\mu\nu}
\end{equation}
with the Einstein tensor $G_{\mu\nu}=R_{\mu\nu}-\frac{1}{2}Rg_{\mu\nu}$, $\kappa=-8\pi G/c^4$, the factor $\sigma(\Psi^i)$ is a coupling to the gravity
while $\Psi^i$ represents other fields, for example
scalar ones. The symmetric tensor $W_{\mu\nu}$ can be any additional geometrical term appearing in the considered ETG. 
The energy-momentum tensor $T_{\mu\nu}$ is the perfect fluid.
We should remember that the tensor $W_{\mu\nu}$ might include extra fields like for instance scalar or electromagnetic ones so besides the modified Einstein's
field equations (\ref{mod1}) we will also deal with equations for the additional fields. For example, as shown in \cite{aw2}, the Klein-Gordon equation of the 
the minimally coupled scalar field with an arbitrary potential plays an important role in the stability analysis.

It has been shown that we may generalize the TOV equations for this class of theories when we introduce the generalization of the energy density and pressure \cite{aw2}:
\begin{align}
 Q(r):=\rho(r)+\frac{\sigma(r)W_{tt}(r)}{\kappa B(r)},\\
\label{def2} \Pi(r):=p(r)+\frac{\sigma(r)W_{rr}(r)}{\kappa A(r)}.
\end{align}
Then, in the case of the spherically symmetric metric
\begin{equation}\label{metric}
 ds^2=-B(r)dt^2+A(r)dr^2+r^2d\theta^2+r^2\sin^2{\theta} d\varphi^2,
\end{equation}
we may write the generalized TOV equations \cite{aw2}
\begin{align}\label{tov}
  \left(\frac{\Pi}{\sigma}\right)'&=-\frac{G\mathcal{M}}{r^2}\left(\frac{Q}{\sigma}+\frac{\Pi}{\sigma}\right)
  \left(1+\frac{4\pi r^3\frac{\Pi}{\sigma}}{\mathcal{M}}\right)\left(1-\frac{2G\mathcal{M}}{r}\right)^{-1}\nonumber\\ 
	&+\frac{2\sigma}{\kappa r}\left(\frac{W_{\theta\theta}}{r^2}-\frac{W_{rr}}{A}\right)\\
	\mathcal{M}(r)&= \int^r_0 4\pi \tilde{r}^2\frac{Q(\tilde{r})}{\sigma(\tilde{r})} d\tilde{r}.\label{mod_mass}
\end{align}
More comments on the above equations can be found in \cite{aw2,anet}. 

We see that the Palatini equations (\ref{EOM_P1}) is of the form (\ref{mod1}) so we may write the TOV equations for Palatini gravity which are
\begin{align}\label{tov_kon}
  \left(\frac{\Pi}{\Phi({r})^2}\right)'&=-\frac{G\tilde{A}\mathcal{M}}{r^2}\left(\frac{Q+\Pi}{\Phi({r})^2}\right)
  \left(1+\frac{4\pi r^3\frac{\Pi}{\Phi({r})^2}}{\mathcal{M}}\right)\nonumber\\ 
	\mathcal{M}(r)&= \int^r_0 4\pi \tilde{r}^2\frac{Q(\tilde{r})}{\Phi(\tilde{r})^2} d\tilde{r},
\end{align}
where $\tilde{A}=\Phi^{-1}A(r)$ and $r$ is the conformal coordinate which should be taken into account in the further analysis. The generalized energy density and pressure are
\begin{subequations}\label{defq}
 \begin{equation}
   \bar{Q}=\bar{\rho}+\frac{1}{2}\bar{U}=\frac{\rho}{\Phi^2}+\frac{U}{2\Phi^2}
 \end{equation}
\begin{equation}
  \bar{\Pi}=\bar{p}-\frac{1}{2}\bar{U}=\frac{p}{\Phi^2}-\frac{U}{2\Phi^2}
\end{equation}
\end{subequations}
with $\bar{U}$ and $\Phi$ depending on the model we are interested in which in our work will be the Starobinsky one. Therefore, 
since $f(\hat{R})=\hat{R}+\beta\hat{R}^2$ and the discussion related to (\ref{EOM_P1}) and (\ref{EOM_scalar_field_P1}),
we get form the structural equation that $\hat{R}=-\kappa T=\kappa c^2\rho$ so the scalar field and the potential appearing in the mentioned field equations are
$\Phi=1+2\kappa c^2\beta\rho$ and $U=\beta\rho^2$. We have already used the discussed in the next subsection \ref{LES} fact that in the case of non-relativistic stars we 
may approximate that
$p<<c^2\rho$ and hence $T=3p-c^2\rho\sim-c^2\rho$.

\subsection{Modified Lane-Emden equation for the Starobinsky $f(\hat{R})=\hat{R}+\beta\hat{R}^2$ Lagrangian}\label{LES}
We are interested in an astrophysical object with the spherical symmetric distribution of matter which can be considered as a useful toy model of a non-relativistic stars, 
like white dwarfs for instance.
As already shown in \cite{anet}, one may consider stable stars' systems in the framework of Palatini gravity, even in the case of the polytropic equation of state 
\cite{olmo_ns,fatibene}
\begin{equation}
 p=K\rho^\gamma,
\end{equation}
with $K$ and $\gamma$ being the parameters of the polytropic EoS. The key observation is that for small values of $p$ the conformal transformation (\ref{met})
preserves the polytropic equation of state \cite{fatibene}. Due to that fact, in the case of the Starobinsky
model $f(\hat{R})=\hat{R}+\beta\hat{R}^2$ we may write:
\begin{align}
 \bar{Q}&=\bar{\rho}+\frac{\bar{U}}{2\kappa}=\frac{4\rho+\beta\kappa^2\rho^2}{4(1+\beta\kappa^2\rho)^2}\sim \rho,\\
 \bar{\Pi}&=\bar{p}-\frac{\bar{U}}{2\kappa}=\frac{4K\rho^\gamma-\beta\kappa^2\rho^2}{4(1+\beta\kappa^2\rho)^2}\sim K\rho^\gamma.
 \end{align}
Bearing it into the mind, we directly end up with the Newtonian equation for (\ref{tov_kon}) assuming, like in General Relativity, that $p<<\rho$ together with
$4\pi r^3 p<< \mathcal{M}$ and $\frac{2G\mathcal{M}}{r}<<1$ which provides
\begin{equation}\label{part_EL}
 -r^2\Phi(r)p'=G\mathcal{M}(r)\rho(r)
\end{equation}
while 
\begin{equation}
 \mathcal{M}(r)= \int^r_0 4\pi \tilde{r}^2\frac{Q(\tilde{r})}{\Phi(\tilde{r})^2} d\tilde{r}\sim \int^r_0 4\pi\rho \tilde{r}^2d\tilde{r}.
\end{equation}
Dividing (\ref{part_EL}) by $\rho$ and differentiating the above two equations with respect to $r$ give us
\begin{equation}
 \frac{d}{dr}\left(\Phi(r)\frac{r^2}{\rho(r)}\frac{dp(r)}{dr}\right)=-4\pi Gr^2\rho(r).
\end{equation}
Using the standard definitions of dimensionless variables
\begin{align}
 r&=r_c\bar{\xi},\;\;\;\rho=\rho_c\theta^n,\;\;\;p=p_c\theta^{n+1},\label{def1}\\
 r^2_c&=\frac{(n+1)p_c}{4\pi G\rho^2_c},\label{def2}
\end{align}
where $p_c$ and $\rho_c$ are the central pressures and densities while $n=\frac{1}{\gamma-1}$ is the polytropic index, we write the modified Lane-Emden equation in the Einstein frame
\begin{equation}
\frac{1}{\bar{\xi}^2}\frac{d}{d\bar{\xi}}\left( \bar{\xi}^2\frac{d}{d\bar{\xi}}\theta+2\alpha\bar{\xi}^2\theta^n\frac{d}{d\bar{\xi}}\theta \right)=-\theta^n. 
\end{equation}
We have already used the exact form of the conformal function:
\begin{equation}\label{fi}
 \Phi=1+2\kappa c^2\beta\rho=1+2\alpha\theta^n
\end{equation}
where $\alpha:=\kappa c^2\beta\rho_c$. The equation can be further written in the more compact way
\begin{equation}
\frac{1}{\bar{\xi}}\frac{d^2}{d\bar{\xi}^2}\left(\left[1+\frac{2\alpha}{n+1}\theta^n\right]\bar{\xi}\theta\right)=-\theta^n. 
\end{equation}
As mentioned above, the equation should be now transformed to Jordan frame: the dimensionless parameter $\bar{\xi}$ comes from the conformal radius which is not a scaled 
value. Thus, we are dealing with the transformation $\bar{\xi}^2=\Phi\xi^2$ such that
\begin{equation}\label{almost}
\xi^2\theta^n\Phi+\displaystyle\frac{\Phi^{-1/2}}{1+\frac12 \xi\Phi_{\xi}/\Phi}\displaystyle\frac{d}{d\xi}
\left(\displaystyle\frac{\xi^2\Phi^{3/2}}{1+\frac12 \xi\Phi_{\xi}/\Phi}\displaystyle\frac{d\theta}{d\xi} \right)=0.
\end{equation}
Let us notice that without assuming any value on the theory parameter $\beta$, the quantity $\alpha$ is already small: for example, for
central density of the order $10^{18}\frac{kg}{m^3}$ we have
\begin{equation}
 \alpha\sim 10^{-7}\beta.
\end{equation}
Due to this fact, in the following calculations we will take into account only the terms which are linear in $\alpha$. Thus, the expansion
of (\ref{almost}) around $\alpha=0$ up to the first order leaves just (after coming back to the more familiar form)
\begin{equation}\label{em}
 \frac{1}{\xi^2}\frac{d}{d\xi}\left[(1+2\alpha\theta^n)\xi^2\frac{d\theta}{d\xi}+\alpha\xi^3\theta^{2n}\right]=-\theta^n+3\alpha\theta^{2n}.
\end{equation}
This is the modified Lane-Emden equation coming from Palatini gravity with Starobinsky term which reduces to the standard one when $\alpha=0$. We are
going to examine it for different types of stars and compare the results to the GR case.

\subsection{Solutions of the modified Lane-Emden equation}

Let us briefly discuss the result which are obtained by solving numerically the modified Lane-Emden equation (\ref{em}). We have considered the polytropic stars with the index
$n=1$, $n=1.5$, and $n=3$, which provide star models for neutron stars, approximation to completely convective stars (Red Giants and Brown Dwarfs), and main sequence stars 
such as Sun and White Dwarfs, respectively. 
It should be mentoned that the equation (\ref{em}) possesses an exact solution for the case $n=0$ (incompressible stars):
\begin{equation}
 \theta(\xi)=1-\frac{\xi^2}{6(1+2\alpha)}
\end{equation}
which recovers the GR solution of the standard Lane-Emden equation when $\alpha=0$. The plot of this solution with respect to the changes of the parameter $\alpha$ is 
presented in the Fig.\ref{fig.4}.

\begin{figure}[h!t]
\centering
\includegraphics[scale=.9]{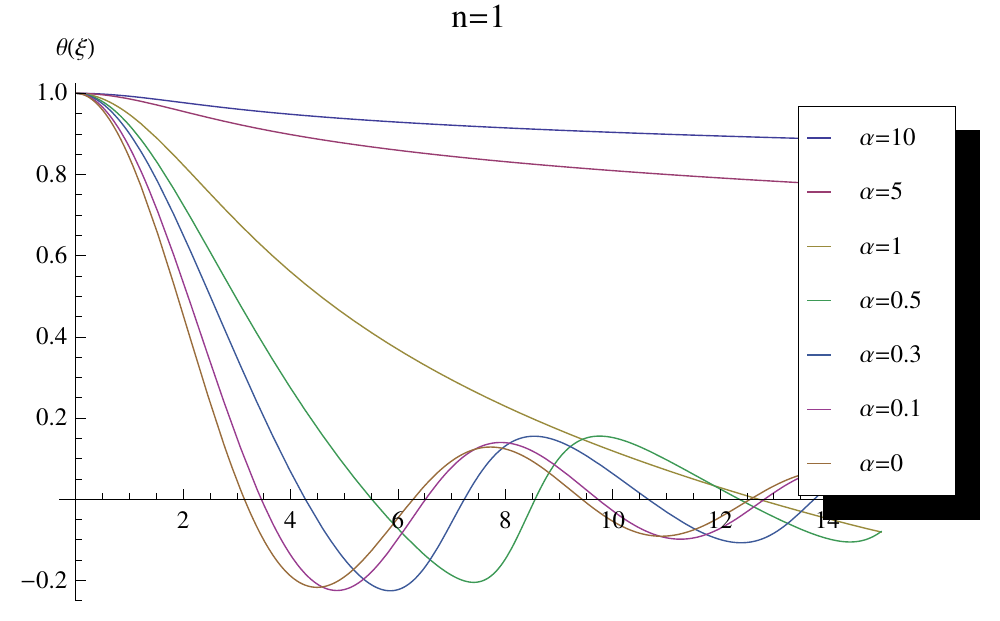}
\includegraphics[scale=.9]{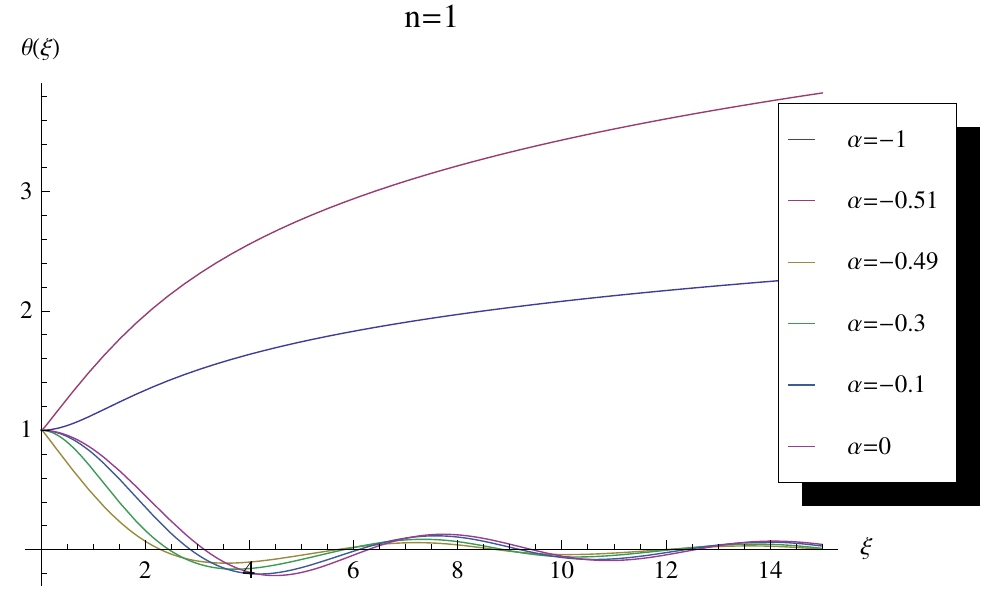}
\caption{(color online) Plots of the numerical solutions for the case $n=1$ for positive (top) and negative (bottom) values of the parameter $\alpha$.}
\label{fig.1}
\end{figure}

\begin{figure}[h!t]
\centering
\includegraphics[scale=.9]{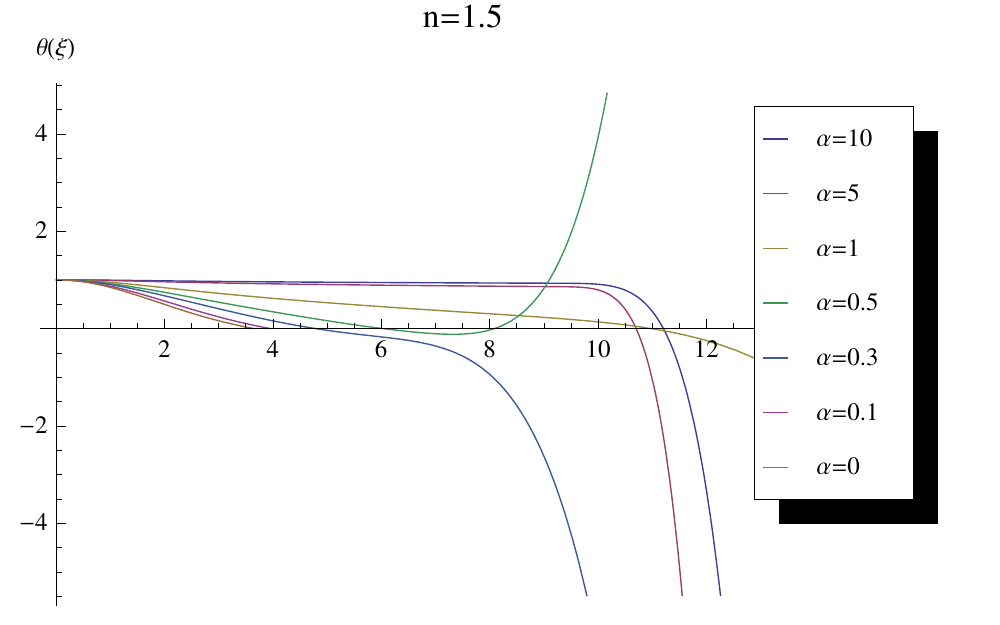}
\includegraphics[scale=.9]{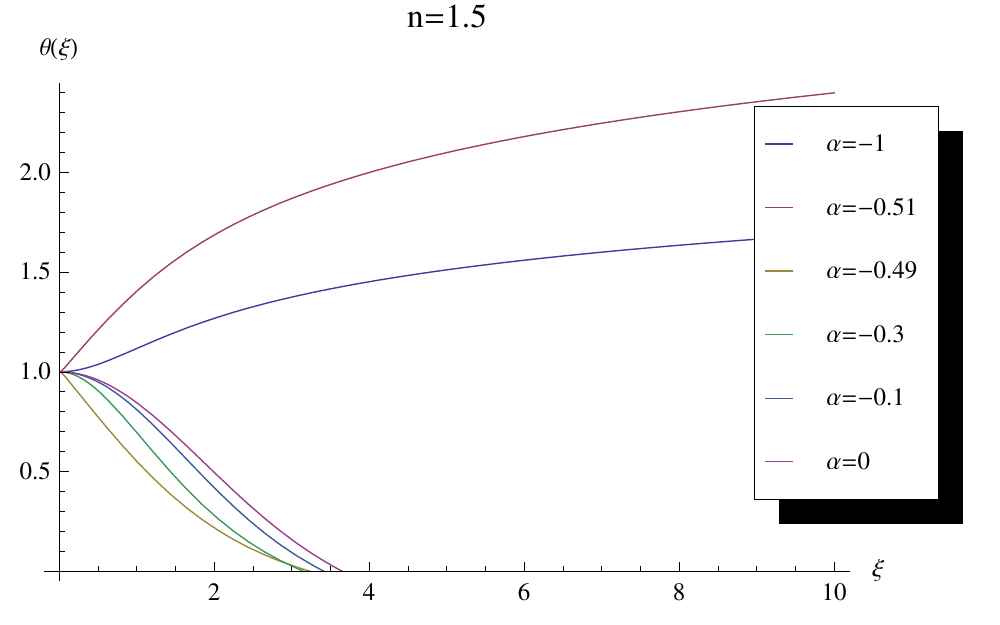}
\caption{(color online) Plots of the numerical solutions for the case $n=1.5$ for positive (top) and negative (bottom) values of the parameter $\alpha$.}
\label{fig.2}
\end{figure}

\begin{figure}[h!t]
\centering
\includegraphics[scale=.9]{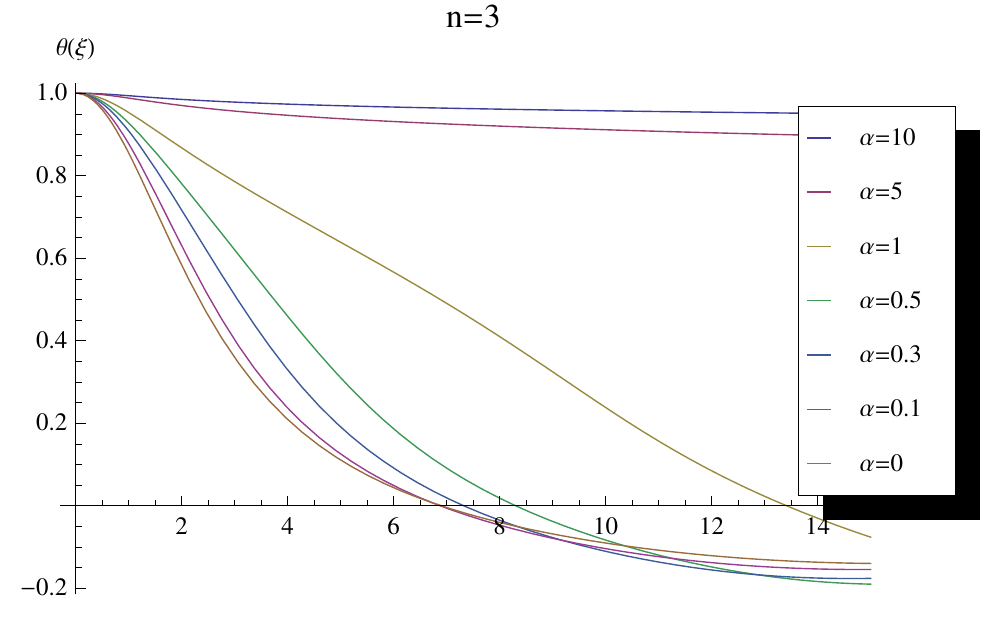}
\includegraphics[scale=.9]{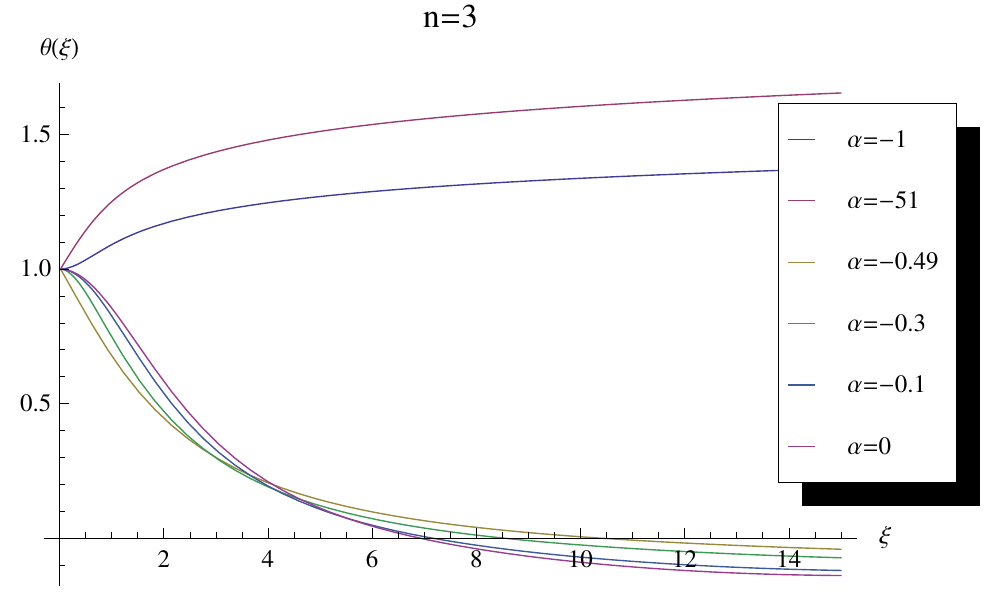}
\caption{(color online) Plots of the numerical solutions for the case $n=3$ for positive (top) and negative (bottom) values of the parameter $\alpha$.}
\label{fig.3}
\end{figure}

\begin{figure}[h!t]
\centering
\includegraphics[scale=.9]{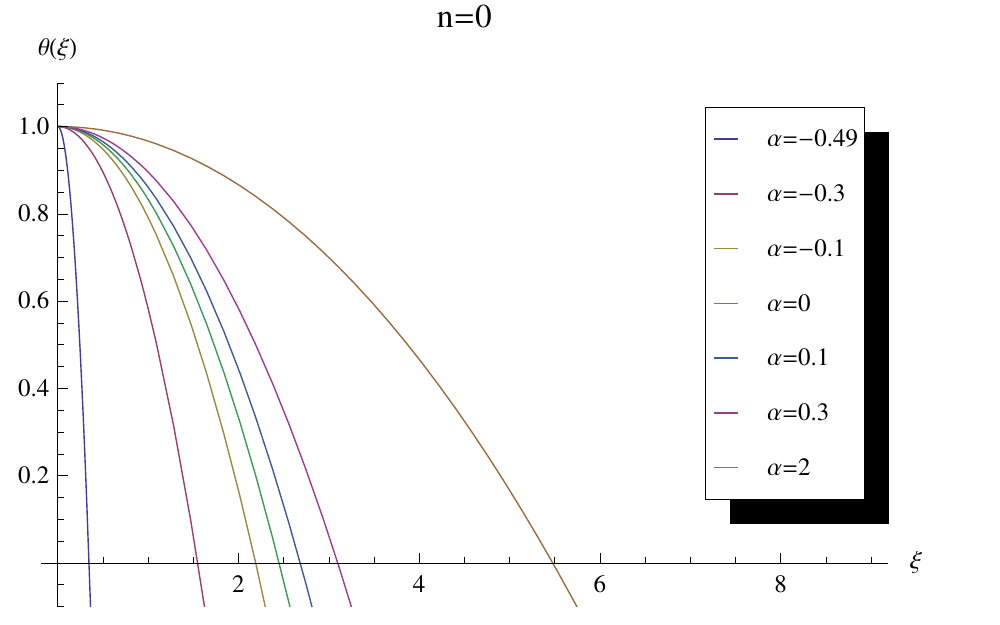}
\caption{(color online) Plots of the exact solutions for the case $n=0$ for a few values of the parameter $\alpha$.}
\label{fig.4}
\end{figure}

In the Fig.\ref{fig.1}, Fig.\ref{fig.2} and Fig.\ref{fig.3} we have plotted the solutions for some values of the parameter $\alpha$. The first observation that we have detected 
is that the bigger positive values of $\alpha$, the larger radius $\xi$ is than the one provided by standard Lane-Emden equation, i.e. the Newtonian 
limit of GR  ($\alpha=0$). There is also no bound on the positive 
$\alpha$: increasing the parameter one approaches $\theta=1$ (the radius is becoming infinite) while decreasing the values of $\alpha$ causes overlapping with the GR curve. The situations differs totally 
in the case of negative parameter: from the numerical analysis we have obtained that there is a bound on the parameter which is $\alpha=-\frac{1}{2}$. This is related to the singular
behavior of the conformal factor (\ref{fi}) for Starobinsky model: let us notice, that for the central density $\rho_c:=\rho(0)$ the function $\theta(0)=1$. Therefore,
there we have
\begin{equation}
 \Phi=1+2\alpha.
\end{equation}
The conformal transformation is singular if the conformal function changes a sign, i.e., $\Phi=0$ and hence $\alpha\neq\frac{1}{2}$.
Below that value we 
do not obtain the physically sensible profiles which is visible in the pictures. Thus, we are left with $\alpha>-\frac{1}{2}$.

It would be also interesting to put some constraint on the upper bound. Since we are dealing with modofications to Newtonian gravity, we work with a weak 
gravitational field. Thus,
 the matter moves slowly and its velocity relative to Solar System center mass is $v^2\leq10^{-7}$. An expression for the circular velocity of an object moving around 
 a mass centre in the case of Starobinsky model in Palatini formalism under EPS interpretation was obtained in \cite{wojn_galax}. 
 Using this expression one may try to bound the parameter $\alpha$ and hence
 considering an object 
 moving around the Sun on the roughly circular orbit with $r=1\text{AU}$ one finds that the upper bound is around $10^4$.

The radius and mass of a star are determined by the first zeros $\xi_1$ of the function $\theta$, that is, $\theta(\xi_1)=0$. The radius is provided 
by (\ref{def1}) and (\ref{def2}) while the mass is given as
\begin{equation}\label{mas}
 \mathcal{M}=4\pi r^3_c\rho_c\int^{\xi_1}_0\xi^2 \theta(\xi)^n d\xi.
\end{equation}
Unfortunately, the modified Lane-Emden equation (\ref{em}) includes the quadratic term in $\theta^n$. It makes obtaining the mass values not so easily as in the Newtonian case 
or modified equations whose the right-hand side looks like the standard one, e.g. \cite{saito,andre}. Since it is multiplied by the small parameter $\alpha$, we will suppress that term in order to have
\begin{equation}\label{simMass}
 \mathcal{M}\sim \xi_1^2|\theta'(\xi_1)|.
\end{equation}

Now on, using the above expression, we compare the quantitative difference for radius and mass of a star obtained from Palatini gravity and GR. Therefore, 
 we have written down the ratios of $R_i/R_0$ and $\mathcal{M}_i/\mathcal{M}_0$ in the 
table \ref{tab}. The values with the index zero are the values for $\alpha=0$ while $i=\{1,3\}$ denotes the value for $n=1$ and $n=3$, respectively. We have not included the 
values obtained for $n=1.5$ because the first zeros in that case were obtained after the extrapolation procedure. Nonetheless, if we look at the figure (\ref{fig.2}), we see 
that for $\alpha>0$ the radius grows together with increasing the parameter. It also happens for the negative values, but the curves' behavior 
starts to resemble the one from the case $n=3$ when we approach the zeros, that is, around $\theta=0$. 

Let us also recall (see e.g. \cite{weinberg}) that one may find the formula relating mass and radius of the star. However, in this 
particular case, because $\theta(\xi_1)=0$, the only difference appears in the value of $\xi_1^2|\theta'(\xi_1)|$ which depends on 
the values of the parameter $\alpha$ and solutions of (\ref{em}) at $\xi=\xi_1$, as written down in the table \ref{tab}. Thus, also in our case the mass (\ref{mas}) as a function of 
a radius is presented as
\begin{align}
 \mathcal{M}(R)=-4\pi R^{\frac{3-n}{n-1}}\left(\frac{(n+1)K}{4\pi G}\right)^\frac{n}{n-1}\xi_1^\frac{n+1}{n-1}\theta(\xi_1)
\end{align}
where we have used $R=r_c\xi_1$ in order to eliminate $\rho_c$ and $K=p_c\rho_c^{-\frac{5}{3}}$. Immediately one notice that 
$\mathcal{M}$ is a constant value for $n=3$ while for $n=3/2$ the relation is analogues to the one in GR, that is,
\begin{align}
\mathcal{M}(R)=-4\pi R^{-3}\left(\frac{5K}{8\pi G}\right)^3\xi_1^5\theta'(\xi_1).
\end{align}

\begin{table}
\centering
\caption{Ratios of numerical values of the radius and masses for the different types of stars (for example $R_i,\;i=\{1,3\}$ denotes a radius ratio for $n=1$, $n=3$, respectively)
for various values of $\alpha$.}
\label{tab}
\begin{tabular}{|c||c c|c c|}
\hline
$\alpha$ & $R_1$ & $M_1$ & $R_3$  & $M_3$  \\
\hline\hline
1 & 4.05 & 2.01  & 1.94 & 4.79   \\
0.5  & 1.75 & 1.52  &  1.2 & 2.02   \\
0.3  & 1.36 & 1.37 &  1.06 & 1.5   \\
0.1  & 1.1 & 1.21  &  0.99 & 1.14   \\
\hline
0    & 1 & 1 &  1 & 1   \\
\hline
-0.1   & 0.91 & 0.85 &  1.03 & 0.88   \\
-0.3   & 0.78 & 0.56 &  1.23 & 0.73   \\
-0.49  & 0.73 & 0.35 &  1.5 & 0.69   \\
\hline
\end{tabular}
\end{table}

\section{Conclusions}
In contrast to the existing works on Palatini stars, we have used the EPS interpretation of the theory which provides different TOV equations. Together with the previous 
studies on neutron stars \cite{anet}, galaxy rotation curves \cite{wojn_galax}, and cosmology \cite{mauro, fatibene1, fatibene2} the current 
proposal has added new arguments in favor of Palatini gravity under the EPS formulation.

We have derived the Lane-Emden equation coming from the Palatini modified equations describing the relativistic stellar object. Apart from the quadratic term in $\theta^n$ on the 
right-hand side it resembles the modified equations obtained already in the literature \cite{riazi, capH, saito, andre, sak, koy}. The numerical solutions of the equation 
(\ref{em}) pictured in the figures (\ref{fig.1}), (\ref{fig.2}) and (\ref{fig.3}) 
definitely shows that we deal with larger stars 
together with increasing the parameter $\alpha$ for $n=1$. The case $n=3$ differs a lot: for the positive parameter $\alpha$ the situation is similar like for $n=1$ while 
for negative values (see the curves in the bottom of the figure (\ref{fig.3})) is opposite in the case of the radius: the star is larger with respect to decreasing $\alpha$ while 
masses tend to decrease.
Moreover, independently of the type of a star, because of the conformal 
transformation the case $\alpha=-\frac{1}{2}$ is excluded while below $\alpha=-\frac{1}{2}$ there are unphysical profiles.

The masses are significantly larger than in GR case, especially for bigger values of the parameter in both cases. Decreasing $\alpha$ one obtains smaller masses where in 
the case of negative values of the parameter we deal with masses smaller than the ones predicted from GR. We have not considered the masses for the case $n=1.5$ 
because the numerical solutions suffered by the extrapolation procedure and we do not treat the results reliable. We should also remember that in all cases we have used the 
simplified mass formula (\ref{simMass}).

Although our studies should be viewed as a toy model, we consider it as a first step to the more accurate stellar description which we leave for the future projects. In this sense, the recent finding of a mapping between Palatini theories of gravity and GR \cite{Afonso:2018bpv,Afonso:2018mxn} may be helpful for this analysis. Work along these lines is currently underway.




\section*{Acknowledgements}
The author would like to thank Gonzalo Olmo, Diego Rubiera-Garcia, Artur Sergyeyev, and Hermano Velten for their comments and helpful discussions.
The work is supported by the NCN grant $DEC-2014/15/B/ST2/00089$ (Poland) and FAPES (Brazil). 



\end{document}